\RequirePackage{fix-cm}
\documentclass[smallextended]{svjour3}       
\smartqed  

\usepackage{graphicx}
\usepackage{amsmath}
\usepackage{amssymb}
\usepackage{multirow}
\usepackage{hhline}
\usepackage[ruled]{algorithm2e}
\usepackage{indentfirst}

\newtheorem{statement}{Statement}

\SetKwProg{Fn}{Function}{:}{}
\DontPrintSemicolon

\begin{document}

\title{Efficient Online Sensitivity Analysis For The Injective Bottleneck Path Problem
\thanks{The work of the author Malyshev D.S. was conducted within the framework of the Basic Research Program at the National Research University Higher School of Economics (HSE).}}


\titlerunning{Efficient Online Sensitivity Analysis$\ldots$}        

\author{Kirill V. Kaymakov, Dmitry S. Malyshev}

\authorrunning{K. V. Kaymakov, D. S. Malyshev} 

\institute{
K. V. Kaymakov \at
              Coleman Tech LLC,\\
              40 Mira Avenue, Moscow, 129090, Russia. \\
              Tel.:+7 (495) 640-17-55\\
              \email{kirill.kaymakov@mail.ru}
\and
D. S. Malyshev \at
              Laboratory of Algorithms and Technologies for Networks Analysis,\\
              National Research University Higher School of Economics,\\
              136 Rodionova Str., Nizhny Novgorod, 603093, Russia. \\
              Tel.:+7 (831) 436-13-97\\
              \email{dsmalyshev@rambler.ru}
}

\date{Received: date / Accepted: date}

\maketitle

\begin{abstract}
The tolerance of an element of a combinatorial optimization problem with respect to a given optimal solution is the maximum change, i.e., decrease or increase, of its cost, such that this solution remains optimal. The bottleneck path problem, for given an edge-capacitated graph, a source, and a target, is to find the $\max$-$\min$ value of edge capacities on paths between the source and the target. For any given sample of this problem with $n$ vertices and $m$ edges, there is known the Ramaswamy-Orlin-Chakravarty's algorithm to compute an optimal path and all tolerances with respect to it in $O(m+n\log n)$ time. In this paper, for any in advance given $(n,m)$-network with distinct edge capacities and $k$ source-target pairs, we propose an $O\Big(m \alpha(m,n)+\min\big((n+k)\log n,km\big)\Big)$-time preprocessing, where $\alpha(\cdot,\cdot)$ is the inverse Ackermann function, to find in $O(k)$ time all $2k$ tolerances of an arbitrary edge with respect to some $\max\min$ paths between the paired sources and targets. To find both tolerances of all edges with respect to those optimal paths,
it asymptotically improves, for some $n,m,k$, the Ramaswamy-Orlin-Chakravarty's complexity $O\big(k(m+n\log n)\big)$ up to $O(m\alpha(n,m)+km)$.

\end{abstract}

\section{Introduction}

First of all, we present known definitions and results, concerning combinatorial optimization problems and their sensitivity analysis, following definitions and notations from \cite{GJM06,TJ22}. A \emph{combinatorial minimization problem} is a problem of selecting a group of elements from a ground set, such that this combination yields a feasible solution and has the lowest cost among all feasible solutions. CMPs can have different objectives. The \emph{sum objective function} minimizes the sum of the costs, whenever the \emph{bottleneck} (or $\min$-$\max$) \emph{objective function} minimizes the maximum cost among all elements in the solution. These CMPs are called \emph{additive} and \emph{bottleneck} (or $\min$-$\max$), respectively. There are many concrete additive and bottleneck CMPs, see, for example, \cite{ABV09,ACDR02,BDM09,CGR96,KPP04}.

More formally, following \cite{TJ22}, an instance of a CMP is given by a tuple $({\mathcal E},F,c,f_c)$, where ${\mathcal E}$ is the finite \emph{ground set of elements}, $F\subseteq 2^{\mathcal E}\setminus \{\emptyset\}$ is the \emph{set of feasible solutions}, $c: {\mathcal E} \longrightarrow {\mathbb R}$ is the \emph{cost function}, $f_c: F \longrightarrow {\mathbb R}$ is the \emph{objective function}. A feasible solution $S^*\in F$ is called an \emph{optimal solution} of the CMP if $f_c(S^*)=\min\limits_{S\in F}f_c(S)$.
The function $f_c(\cdot)$ is the \emph{sum objective function} if $f_c(S)=\sum\limits_{e\in S}c(e)$, and it is the \emph{bottleneck objective function} if $f_c(S)=\max\limits_{e\in S} c(e)$. Note that CMPs with these objectives are equivalent to the maximization ones, more precisely, to a $\max$-$\min$ problem if those CMP is of the $\min$-$\max$ type, by taking the costs with the opposite sign.

Sensitivity analysis is the field of finding limit cost changes, for which an optimal solution remains optimal. More specifically, upper tolerances measure the supremum increase in the cost of an element, such that a current solution remains optimal, and lower tolerances measure the corresponding supremum decrease \cite{GJM06}.

If $\Pi=({\mathcal E},F,c,f_c)$ is an instance of a combinatorial minimization or maximization problem, $e\in {\mathcal E}$, and $\alpha\in \mathbb{R}$ is some constant, then, by $\Pi_{e,\alpha}=({\mathcal E},F,c_{e,\alpha},f_{c_{e,\alpha}})$, a problem is denoted with

$$c_{e,\alpha}(e)=c(e)+\alpha, c_{e,\alpha}(e')=c(e')\,\, \forall e'\neq e.$$
Let $S^*$ be an optimal solution of $\Pi$ and $e\in {\mathcal E}$. The \emph{upper tolerance of $e$ $($with respect to $S^*)$} is defined as

$$u_{S^*}(e)=\sup\{\alpha \in \mathbb{R}_{\geq 0}:~S^*~\text{is an optimal solution of}~\Pi_{e,\alpha}\}.$$

The \emph{lower tolerance of $e$ $($with respect to $S^*)$} is defined as

$$l_{S^*}(e)=\sup\{\alpha \in \mathbb{R}_{\geq 0}:~S^*~\text{is an optimal solution of}~\Pi_{e,-\alpha}\}.$$

By $F_{+e}$ and $F_{-e}$, we denote the sets of those feasible solutions of $\Pi$ that
contain and do not contain $e$, respectively, i.e.
$$F_{+e}=\{S\in F:~e\in S\},\,\, F_{-e}=\{S\in F:~e\not \in S\}.$$
By $f_c(F),f_c(F_{+e})$, and $f_c(F_{-e})$, we denote the optimal values of the objective function on $F,F_{+e},F_{-e}$, i.e.
$$f_c(F)=\min\limits_{S\in F} f_c(S),\,\,f_c(F_{+e})=\min\limits_{S\in F_{+e}} f_c(S),\,\,f_c(F_{-e})=\min\limits_{S\in F_{-e}} f_c(S)~\text{or}$$
$$f_c(F)=\max\limits_{S\in F} f_c(S),\,\,f_c(F_{+e})=\max\limits_{S\in F_{+e}} f_c(S),\,\,f_c(F_{-e})=\max\limits_{S\in F_{-e}} f_c(S).$$

The following statement has been proven in \cite{GJM06}, see Theorems 4 and 15 of those paper:

\begin{statement}
\label{st1}
Let $\Pi=({\mathcal E},F,c,f_c)$ be an instance of a combinatorial minimization/maximization problem, $S^*$ be an optimal solution of $\Pi$, $e\in {\mathcal E}$ be an element of the ground set. If $\Pi$ is an additive minimization problem, then

$$u_{S^*}(e)=f_c(F_{-e})-f_c(F),~\text{if}~e\in S^*,\,\,u_{S^*}(e)=+\infty,~\text{if}~e\in {\mathcal E}\setminus S^*;$$
$$l_{S^*}(e)=f_c(F_{+e})-f_c(F),~\text{if}~e\in {\mathcal E}\setminus S^*,\,\,l_{S^*}(e)=+\infty,~\text{if}~e\in S^*.$$

If $\Pi$ is a $\min$-$\max$ problem, then

$$u_{S^*}(e)=f_c(F_{-e})-c(e),~\text{if}~e\in S^*,\,\,u_{S^*}(e)=+\infty,~\text{if}~e\in {\mathcal E}\setminus S^*;$$
$$l_{S^*}(e)=c(e)-f_c(F),~\text{if}~e\in {\mathcal E}\setminus S^*~\text{and}~\min\limits_{S\in F_{+e}}\max\limits_{e'\in S\setminus \{e\}}c(e')<c(e),$$
$$l_{S^*}(e)=+\infty,~\text{otherwise}.$$

If $\Pi$ is an additive maximization problem, then

$$u_{S^*}(e)=f_c(F)-f_c(F_{+e}),~\text{if}~e\in {\mathcal E}\setminus S^*,\,\,u_{S^*}(e)=+\infty,~\text{if}~e\in S^*;$$
$$l_{S^*}(e)=f_c(F)-f_c(F_{-e}),~\text{if}~e\in S^*,\,\,l_{S^*}(e)=+\infty,~\text{if}~e\in {\mathcal E}\setminus S^*.$$

If $\Pi$ is a $\max$-$\min$ problem, then

$$u_{S^*}(e)=f_c(F)-c(e),~\text{if}~e\in {\mathcal E}\setminus S^*~\text{and}~\max\limits_{S\in F_{+e}}\min\limits_{e'\in S\setminus \{e\}}c(e')>c(e),$$
$$u_{S^*}(e)=+\infty,~\text{otherwise};$$
$$l_{S^*}(e)=c(e)-f_c(F_{-e}),~\text{if}~e\in S^*,\,\,l_{S^*}(e)=+\infty,~\text{if}~e\in {\mathcal E}\setminus S^*.$$
\end{statement}

It follows from Statement \ref{st1} that if a tolerance is finite, then it does not depend on an optimal solution. By the same reason, an element $e\in {\mathcal E}$ belongs to some optimal solution of an additive combinatorial minimization/maximization problem if and only if its lower/upper tolerance is $+\infty$.

Tolerances give to a decision maker information about the stability of an optimal solution with respect to perturbations of its elements. They have also been used to design algorithms for NP-hard and polynomially solvable combinatorial problems, like versions of the Assignment Problem \cite{DT04}, Travelling Salesman Problem \cite{GGT12,TGGS08,TMGP17}, Vehicle Routing Problem \cite{BGKP12}, Weighed Independent Set Problem
\cite{GMPZ15}, see the surveys \cite{TJ22,TMGP17} also. Efficiency of tolerance computations is an important problem, which not only had applications in the mentioned problems, but it is also of independent interest. Several papers have been devoted to this question \cite{BW94,CH78,DRT92,GMP13,JV86,MP14,P15,ROC05,SW80,T82}.

In this paper, we consider sensitivity analysis for the \emph{Bottleneck Path Problem}, abbreviated as the BPP. In this problem, a simple, connected graph $G=(V_G,E_G)$ with $V_G=\{v_1,v_2,\ldots,v_n\}$ and $E_G=\{e_1,e_2,\ldots,e_m\}$ is given and, for every edge $e_i$, its capacity $c_i$ is also given. Additionally given a source vertex $s\in V_G$ and a target vertex $t\in V_G$, the BPP is to find the value $b_G(s,t)=\max\limits_{P\in {\mathcal P}_{st}} \min\limits_{e\in P} c(e)$, where ${\mathcal P}_{st}$ is the set of all paths between $s$ and $t$.
A \emph{path} is a sequence of vertices, in which any two consecutive members form an edge, without repetitions of its vertices and edges.
The BPP arises as a subroutine in several flow problems, see, for example, \cite{BKS02} and \cite{EK72}. The \emph{online} BPP is the problem, when sources $s$ and targets $t$ are entered in the online regime. To the best of our knowledge, this problem was introduced in \cite{KM24}, where it was named the \emph{online Multi-Pair Bottleneck Paths Problem}. To solve the online BPP, an input edge-capacitated graph can be preprocessed in $O(m+n\log n)$ time, such that any bottleneck value $b_G(s,t)$ can be computed in $O(\log n)$ time, see \cite{KM24}.

In this paper, for any in advance given $(G=(V_G,E_G),c)$ with injective $c(\cdot)$ and $(s_1,t_1),\ldots,(s_k,t_k)$, where $s_i,t_i\in V_G$, for any $i$, we present a preprocessing in $O\Big(m \alpha(m,n)+\min\big((n+k)\log n,km\big)\Big)$ time, allowing to compute in $O(k)$ time all $2k$ tolerances of an arbitrary edge with respect to some $\max\min$ $s_it_i$-paths, for all $i$. We assume that the edge capacities are pairwise distinct, and this restriction is used in justification for correctness of our algorithm, see the reasonings before Algorithm \ref{alg3}.

Note that computation of upper and lower tolerances of all edges for the BPP, including determination of an optimal solution, can be done in $O(m+n\log n)$ time \cite{ROC05}. But the approach from \cite{ROC05} does not allow to compute rapidly values of tolerances for individual edges. Analysis of namely individual tolerances may be important, for example, in sustainability research of flows in networks under incidents with a single edges, like breakdowns or repairs of pipelines, possibly, with respect to several source-target pairs. Moreover, to find both tolerances of all edges with respect to some optimal $s_it_i$-paths,
our algorithm asymptotically improves, for some $n,m,k$ (e.g., $m=O(n)$), the Ramaswamy-Orlin-Chakravarty's complexity $O\big(k(m+n\log n)\big)$ up to $O(m\alpha(n,m)+km)$.

We will use the formulae from \cite{T82} on sensitivity analysis for the minimum spanning tree problem and the algorithms from \cite{DRT92,P15} with the complexities $O\big(m\alpha(m,n)\big)$ and $O\big(m\log \alpha(m,n)\big)$, respectively, to compute all tolerances for this problem.
But our algorithm is not completely reduced to those sensitivity analysis, involving some additional arguments. We also propose an algorithm for sensitivity analysis of the minimum spanning tree problem, which is much simpler to understand, than those in \cite{DRT92,P15}, but with the worse complexity $O(m\log n)$.

\section{Our algorithm}

\subsection{Bottleneck paths and maximum spanning trees}
\label{s2.1}

A tree subgraph, containing all vertices of a graph, is called a \emph{spanning tree}. The \emph{maximum spanning tree problem}, abbreviated as the MSTP, is to find a spanning tree with the maximum sum of weights of its edges in a given edge-weighted graph. Several classic algorithms have been designed to solve the MSTP, like the Boruvka's \cite{B26}, the Kruskall's \cite{K56}, the Prim's \cite{P57}, the Chazelle's \cite{Ch00} algorithms. For a graph with $n$ vertices and $m$ edges, the Prim's algorithm can be implemented in $O(m+n\log n)$ time, using Fibonacci heaps \cite{FT87}.
The Chazelle's algorithm is the best known algorithm among deterministic to solve the MSTP, it has the computational complexity $O\big(m\alpha(m,n)\big)$.

The MSTP has an important connection to the BPP, shown in Statement \ref{st2}. This is a well-known fact, but we did not know a corresponding reference, thus, a formal proof was presented in \cite{KM24}.

\begin{statement}
\label{st2}
Let $T$ be an arbitrary maximum spanning tree of $(G,c)$. Then, for any $s\in V_G,t\in V_G$, the minimum capacity of edges  on the path between $s$ and $t$ in $T$ equals $b_G(s,t)$.
\end{statement}

Statement \ref{st2} makes to think of using the following strategy for computing the tolerances: monitor changes of a MST and minima in $st$-paths under capacity changes of individual edges. The Ramaswamy-Orlin-Chakravarty's algorithm explores it, and it will also be applied in our algorithm. Ramaswamy, Orlin, and Chakravarty directly compute $\max\min$ values on $st$-paths in resulting trees, but we involve more sophisticated arguments.

\subsection{Lowest common ancestors and some their applications}
\label{s2.2}

For a given rooted tree and its vertices $x$ and $y$, the \emph{lowest common ancestor} of $x$ and $y$, abbreviated as $LCA(x,y)$, is the deepest node, for which both $x$ and $y$ are descendants, assuming that each vertex is a descendant of itself. It was shown in \cite{FH06} that with a preprocessing step in $O(n)$ time, where $n$ is the vertex number in the tree, LCA of any pair of vertices can be found in $O(1)$ time. The approach from \cite{BF04}, the so-called jump pointers algorithm with preprocessing time $O(n\log n)$, has been modified in \cite{KM24} for computing in logarithmic on $n$ time the minimum and maximum values on in online given paths of an in advance given edge-weighted tree. It can also be easily modified to return optimal edges on these paths.

LCAs are also useful to solve the following task, which will be used to compute the tolerances efficiently. Assume that we are given a tree $T$. For any vertices $x$ and $y$, by $T(x,y)$, we denote the path between $x$ and $y$ in $T$. We need to preprocess $T$ quickly enough, such that, for any given vertices $s,t$ of $T$ and its edge $xy$, checking whether $xy\in T(s,t)$ or not can also be performed rapidly.

It can be done in $O(1)$ time with an $O(n)$ preprocessing stage. To this end, $T$ is rooted at an arbitrary vertex $r$ and it is preprocessed, according to \cite{FH06}. The \emph{depth of a vertex $x\in V(T)$ with respect to} $r$, i.e., $|T(r,x)|$, is denoted by $d_T(x)$. By breadth-first search, we find all depths in $O(n)$ time. Denote by $T_x$ and $T_y$
the connected components of $T\setminus \{xy\}$, containing $x$ and $y$, respectively. Without loss of generality, let us assume that $d_T(y)=d_T(x)+1$. This means that $x$ and $r$ belong to $T_x$. Clearly that
$xy\in T(s,t)$  if and only if $s$ and $t$ lie in distinct components. Checking the fact that a vertex $z\in \{s,t\}$ belongs to $T_y$ can be done by checking the equality $LCA(z,y)=y$. The exactly one of $s$ and $t$ must poses this property to satisfy $xy\in T(s,t)$. It gives an $O(1)$-time algorithm to solve the task above.

\subsection{Disjoint-set data structure}
\label{s2.3}

A \emph{disjoint-set data structure}, DJS, for short, is a data structure
that stores a partition of a finite set into its disjoint subsets. It supports the following operations:

\begin{itemize}
\item $Create(x)$ --- creating the new singleton subset $\{x\}$ and adding it to the structure,

\item $Find(x)$ --- finding a canonical element of those subset,
which contains $x$,

\item $Join(x,y)$ --- replacing the two subsets with the canonical elements $x$ and $y$ by their union
\end{itemize}

in near-constant time. More precisely, insertion and join can be performed in unit time in the worst case, but search can be performed in amortized time,
bounded from above by a value of the inverse Ackermann function, see \cite{TL84}. DJSs are useful in efficient implementation, see, for example, \cite{CLRS22}, of the Kruskal's algorithm, which orders edges by their weights, scans their sorted set, and determines whether a current edge can be added to an optimal solution or not. We will use a similar idea in our algorithm.

\subsection{Efficient sensitivity analysis for the MSTP}
\label{s2.4}

Let $T$ be an arbitrary maximum spanning tree of the graph $(G,c)$. It can be computed in $O\big(m\alpha(m,n)\big)$ time \cite{Ch00}. It was proved, see the paper \cite{T82}, that, for the MSTP on $(G,c)$ and its optimal solution $T$, the following relations are true:

$$u_T(xy)=+\infty, \text{~if~} xy\in E_T;\,\, l_T(xy)=-\infty,\text{~if~} xy\in E_G\setminus E_T;$$
$$u_T(xy)=\min\big\{c(x'y'):~x'y'\in T(x,y)\big\}-c(xy),\text{~if~} xy\in E_G\setminus E_T;$$
$$l_T(xy)=c(xy)-\max\big\{c(x'y'):~xy\in T(x',y')\big\}, \text{~if~} xy\in E_T.$$

Based on these formulae, an algorithm has been designed in \cite{DRT92} for computing all tolerances for the MSTP on $(G,c)$, assuming that an optimal tree has been given. It has the worst-case complexity $O\big(m\alpha(m,n)\big)$, but with $O(m)$ randomized complexity. The result from \cite{DRT92} has been updated to the $O\big(m\log \alpha(m,n)\big)$ complexity \cite{P15}. Together with computing all tolerances, it is possible to determine the following edges in $O\big(m\alpha(m,n)\big)$ or $O\big(m\log \alpha(m,n)\big)$ time:

$$\forall xy\in E_G\setminus E_T\,\,\, \arg\min\limits_{\big\{x'y':~x'y'\in T(x,y)\big\}} c(x'y'),$$
$$\forall xy\in E_T\,\,\, \arg\max\limits_{\big\{x'y':~xy\in T(x',y')\big\}}c(x'y').$$

Indeed, computed the corresponding $\arg\min$s/$\arg\max$s $x'y'$, called the \emph{replacement edges}, the trees $T'=\big(T\setminus \{x'y'\}\big)\cup \{xy\}$ and $T'=\big(T\setminus \{xy\}\big)\cup \{x'y'\}$ will be the maximum spanning trees of $(G,c)$ among its spanning trees, containing $xy\in E_G\setminus E_T$ or not containing $xy\in E_T$, respectively. For any edge of $(G,c)$, its replacement edge is unique if any, as all capacities are pairwise distinct. These observations will be useful for our aims, according to Statement \ref{st2}.

The algorithms from \cite{DRT92,P15} are quite difficult to understand. Here, we present much simpler alternative LCA- and DJS-based algorithms with the worst-case complexities $O(m\log n)$. A pseudo code for determining the replacement edges, corresponding to MSTP upper tolerances, is presented in Algorithm \ref{alg1}. Its computational complexity is $O(m\log n)$, which is obvious.

\begin{algorithm}[h!]
    \caption{\small MSTP upper tolerances replacement edges computation}
    {
    \small
        {\bf Input:} A simple, edge-weighted, connected graph $G=(V_G,E_G,c)$
        and some its maximum spanning tree $T$

        {\bf Output:} An array $U[]$, whose indices are edges of $G$ and elements are the corresponding replacement edges

        \smallskip

        Build the corresponding LCA-based data structure by \cite{KM24};\;

        \For{$(e=xy\in E_G)$}
        {
            \If{$\big(e \in E_T\big)$}
            {
             $U[e] \gets $'No';\;
            }
            \Else
             {
                $U[e] \gets$ find the replacement edge for $e$ by searching the $\arg\min$ on $T(x,y)$;\;
             }
        }
    }
    \KwRet $U[]$
\label{alg1}
\end{algorithm}

Now, let us consider determining the replacement edges, corresponding to MSTP lower tolerances. Let $T$ be any MST of $(G,c)$, rooted at an arbitrary vertex. Sort edges from $E_G\setminus E_T$ by decreasing the capacities. Then, for any $e\in E_T$, its replacement edge is the first edge $xy\in E_G\setminus E_T$ with respect to the order with $e\in E_{T(x,y)}$. Scanning the ordered set $E_G\setminus E_T$, our algorithm will catch the corresponding moments, for all edges in $E_T$.

Let us define a multigraph $(G',c')$, i.e., multiple edges are allowed. For any $xy \in E_G\setminus E_T$, change $xy$ to the edges $xz$ and $yz$, where $z=LCA(x,y)$, with the same capacity $c(xy)$. The tree $T$ is a MST of $(G',c')$, see \cite{DRT92}, and, obviously, any edge of $(G',c')$ connects an ancestor with its descendant. Put $E'=E_{G'}\setminus E_T$.

Together with $T$, we will keep a DJS on $V_T$, whose all elements, i.e., subsets of $V_T$, are vertices of some tree $T'$. Initially, DJS contains all $n$ singletons, corresponding to vertices of $T$, and $T'=T$. At any moment,
all vertices of $T'$ are vertex sets of some subtrees of $T$ and all its edges are exactly edges of $T$, for which replacement edges have not yet been determined. Reading an edge $xy\in E'$, by $Find(x)$ and $Find(y)$, we determine the vertices $X,Y\in V_{T'}$, such that $x\in X$ and $y\in Y$. Suppose that $X\neq Y$. The invariant of the process is that $Y$ is a descendant of $X$, or vice versa, in $T'$, determined by the descendant relation between $x$ and $y$ in $T$. Supposing that $Y$ is a descendant of $X$ in $T$, we contract $T'(X,Y)$ in $T'$ and join the corresponding subsets of $V_T$ into a single vertex in $T'$. Walking through $T'(X,Y)$, for any its edge with a capacity $c^*$, we assign $e$ as the replacement edge of $e'\in E_T$ with $c^{-1}(e')=c^*$.

A pseudo code for determining the replacement edges, corresponding to MSTP lower tolerances, is presented in Algorithm \ref{alg2}. Its computational complexity is, clearly, $O(m\log n)$.

\begin{algorithm}[ht!]
\small
    \caption{\small MSTP lower tolerances replacement edges computation}
    {
        {\bf Input:} A simple, edge-weighted, connected graph $G=(V_G,E_G,c)$ and some its maximum spanning tree $T$

        {\bf Output:} An array $L[]$, whose indices are edges of $G$ and elements are the corresponding replacement edges

        \smallskip

        Build the corresponding data structure by \cite{FH06};\;

        Put $V_{G'} = V_G$;\;

        \For{$(e=xy\in E_G)$}
            {
        $L[e] \gets $'No';\;

        $z\gets LCA(x,y)$;\;

        $E_{G'}\gets E_{G'}\cup \{xz,yz\}$ with $c'(xz)=c'(yz)=c(xy)$;\;
           }

        Put $E'\gets E_{G'}\setminus E_T$ and sort its elements in weights decreasing;\;

        Build the corresponding DJS data structure and $T'$;\;

        \For{$(e=xy\in E')$}
        {
        $X = Find (x), Y = Find (y)$;\;
        \If{$(X\neq Y)$}
           {
           Put $Z\gets Y,Z_{\cup}\gets Y$;\;
           \While{$(Z\neq X)$}
                {
                $c^*\gets c\big(Z,P'(Z)\big)$, where $P'(\cdot)$ is the parenthood relation on $T'$;\;

                $L[e']\gets e$, where $c^{-1}(e')=c^*$;\;

                Join $Z_{\cup}$ and $P'(Z)$ and contract $ZP'(Z)$ in $T'$;\;

                $Z\gets P'(Z)$;\;
               }
            }
        }
    }
    \KwRet $L[]$
\label{alg2}
\end{algorithm}

\subsection{Combining all together}
\label{s2.4}

Suppose that some vertices $s,t$ and an edge $e=xy$ are given in the graph $(G,c)$. Let $T$ be an arbitrary MST of $(G,c)$.
Let $P^*$ be an arbitrary $\max\min$ $st$-path of $(G,c)$ and $e^*$ be its edge $\arg\min\limits_{e\in P^*}c(e )$. The edge
$e^*$ is unique, as $c(\cdot)$ is injective. It is clear that $u_{P^*}(e)=+\infty\,\, \forall e\in E_T$ and $l_{P^*}(e)=+\infty\,\, \forall e \in E_G\setminus E_T$, because the corresponding changes of $c(e)$ do not break the optimality of $T$ as a MST. By the same reason, if a tolerance (upper or lower) of $e$ is finite, then there is a replacement edge $e'$ and this tolerance is at least $|c(e')-c(e)|$. It follows from Statement \ref{st1} that if $u_{P^*}(e)<+\infty$, then $u_{P^*}(e)=c(e^*)-c(e)$.

Let us assume that $e\in T(s,t)$, i.e., $e\in E_T$, which can be checked in constant time, see Subsection \ref{s2.2}. If $L[e]=$'No', i.e., removing $e$ disconnects $(G,c)$, then $l_{P^*}(e) = +\infty$, because any decrease of $c(e)$ keeps the optimality of $T$ for the MSTP and the resultant graph. Suppose that there is an edge $e'=x'y'=L[e]$. It is clear that $e'\not \in E_T$. Then, by Statements \ref{st1} and \ref{st2}, we have $l_{P^*}(e) = c(e) - \min\big(c(e'),c(e^*)\big)$, since
$$\min\limits_{\tilde{e}\in T(x',y')\cup \{e'\}} c(\tilde{e})=c(e'),~\text{where}~T'=\big(T\setminus \{e\}\big)\cup \{e'\},~\text{by the maximality of}~T,$$
$$\min\limits_{\tilde{e}\in T(s,t)} c(\tilde{e})=c(e^*).$$

Suppose that $e=xy\not \in T(s,t)$. Then, $l_{P^*}(e)=+\infty$, as $T(s,t)$ exists in a new tree after any decrease of $c(e)$. If $U[e]=$'No', then \mbox{$u_{P^*}(e) = +\infty$}. Suppose that there is an edge $e'=U[e]$. Then, $e\not \in E_T, e'\in E_T$. If $e'\not \in T(s,t)$, which can be verified in $O(1)$ time, see Subsection \ref{s2.2}, then $u_{P^*}(e) =+\infty$. Indeed, for any increase of $c(e)$, either $T$ or $T'=\big(T\setminus \{e'\}\big)\cup \{e\}$ is an optimal solution of the MSTP. The path $T(s,t)$ exists in both these trees. Suppose that $e'\in T(s,t)$. Then,
$u_{P^*}(e)=c(e^*)-c(e)$ if $c(e')=c(e^*)$, otherwise $u_{P^*}( e)=+\infty$. Indeed, we have
$$u_{P^*}(e)\geq c(e')-c(e)\geq c(e^*)-c(e),~\text{as}~c(e')\geq c(e^*),$$
$$u_{P^*}(e)>c(e^*)-c(e) \longrightarrow u_{P^*}(e)=+\infty,$$
and since $c(\cdot)$ is injective and $e'=\arg\min\limits_{\big\{\tilde{e}:~\tilde{e}\in T(x,y)\big\}}c(
\tilde{e})$ we have
$$c(e')=c(e^*)\longleftrightarrow e'=e^*\longleftrightarrow e^*\not \in T'( s,t),$$
$$e^*\in T'(s,t)\longrightarrow \arg\min\limits_{\tilde{e}\in T'(s,t)} c(\tilde{e})\in \{e,e^*\}\longrightarrow u_{P^*}(e)=+\infty.$$

A pseudo code of an algorithm for working with a pair of a source and a target is presented in Algorithm \ref{alg3}. Its preprocessing stage, emphasized with the underlines, takes $O\big(m\alpha(m,n)\big)$ time. Its running time is $O(1)$ per an edge.

\begin{algorithm}[h!]
\small
    \caption{\small BPP online tolerances computation}
    {
        {\bf Input:} A simple, edge-capacitated, connected graph $G=(V_G,E_G,c)$, vertices $s,t\in V$ (in advance) and an edge $e=xy\in E$ (in online)

        {\bf Output:} The lower and upper tolerances of $e$ with respect to a $\max\min$ path between $s$ and $t$

        \smallskip

         \underline{Find a maximum spanning tree $T$ of $G$ by \cite{Ch00}};\;

         \underline{Find a $\max\min$ $st$-path $P^*$ and its edge $e^*=\arg\min\limits_{e\in P^*}c(e)$};\;

         \underline{Build the corresponding data structure by \cite{FH06}};\;

         \underline{$(U[],L[])\gets$ apply the algorithm from \cite{P15}};\;

         \If{$(e\in E_{T(s,t)})$}
            {
             $u_{P^*}(e) \gets +\infty$;\;

             \If{$(L[e]=$'No'$)$}
               {
             $l_{P^*}(e) \gets +\infty$;\;
               }
            \Else
              {
             $e'\gets L[e]$;\;

             $l_{P^*}(e) \gets c(e) - \min\big(c(e'),c(e^*)\big)$;\;
               }
            }

        \Else
             {
                $l_{P^*}(e) \gets +\infty$;\;

                \If{$(U[e]=$'No'$)$}
                    {
                    $u_{P^*}(e) \gets +\infty$;\;
                    }
                \Else
                {
                $e'\gets U[e]$;\;

                \If{$\big(e'\not \in T(s,t) \vee c(e')\neq c(e^*)\big)$}
                   {
                   $u_{P^*}(e) \gets +\infty$;\;
                   }

                \Else
                  {
                   $u_{P^*}(e)\gets c(e^*) - c(e)$;\;
                  }
                }
            }
    }

    \KwRet $l_{P^*}[e],u_{P^*}[e]$
\label{alg3}
\end{algorithm}

Algorithm \ref{alg3} can be modified to work with $k$ pairs $(s_1,t_1),\ldots,(s_k,t_k)$ of sources and targets.
To this end, we find all $e^*_i=\arg\min\limits_{e\in P^*_i}c(e)$, where $P^*_i$ is a $\max\min$ $s_it_i$-path, either in $O(km)$ time or in $O(n\log n + k\log n)$ time, using LCA-based approach from Subsection \ref{s2.2}. Hence, all $2k$ tolerances of any given edge with respect to $P^*_1$--$P^*_k$ can
be computed in $O(k)$ time under $O\Big(m \alpha(m,n)+\min\big((n+k)\log n,km\big)\Big)$-time preprocessing. It gives
an $O(m \alpha(m,n)+km)$-time algorithm for computing both tolerances of all edges with respect to $P^*_1$--$P^*_k$,
sometimes improving the Ramaswamy-Orlin-Chakravarty's complexity $O\big(k(m+n\log n)\big)$, e.g., when $k=1$ and $m=O(n)$.

\section{Conclusion and future work}
\label{s3}

In this paper, we considered the bottleneck path and
sensitivity analysis problems in the form of tolerances computation for individual edges with respect to an optimal solution. The previous state-of-the-art algorithm, due to Ramaswamy, Orlin, and Chakravarty, computes an optimal solution and tolerances with respect to it in $O(m+n\log n)$ time. In this paper, for any in advance given distinct-capacities network and $k$ source-target pairs, we propose an $O\Big(m \alpha(m,n)+\min\big((n+k)\log n,km\big)\Big)$-time preprocessing to find in $O(k)$ time all $2k$ tolerances of an arbitrary edge with respect to some $\max\min$ paths between the paired sources and targets. To compute both tolerances of all edges with respect to those optimal paths, it asymptotically improves, for some $n,m,k$, the Ramaswamy-Orlin-Chakravarty's complexity $O\big(k(m+n\log n)\big)$ up to $O(m\alpha(n,m)+km)$. Developing new algorithms and improving existing ones is a challenging research problem for future work.

\section*{Statements and Declarations}

\subsection{Funding}

The work of the author Malyshev D.S. was conducted within the framework of the Basic Research Program at the National Research University Higher School of Economics (HSE).

\subsection{Authors' contribution}

All authors contributed to the study conception and design. Material preparation was performed by Kirill Kaymakov and Dmitriy Malyshev. The first draft of the manuscript was written by Kirill Kaymakov and Dmitriy Malyshev and all authors commented on previous versions of the manuscript. All authors read and approved the final manuscript.

\subsection{Data availability}

Our paper has no associated data.

\end{document}